# The Infrared Imaging Spectrograph (IRIS) for TMT: Instrument Overview

Anna M. Moore[*a], James E. Larkin[b], Shelley A. Wright[c,d], Brian Bauman[e], Jennifer Dunn[f], Brent Ellerbroek[g], Andrew C. Phillips[h], Luc Simard[i], Ryuji Suzuki[j], Kai Zhang[k], Ted Aliado[b], George Brims[b], John Canfield[b], Shaojie Chen[c], Richard Dekany[a], Alex Delacroix[a], Tuan Do[c,d], Glen Herriot[f], Bungo Ikenoue[j], Chris Johnson[b], Elliot Meyer[c,d], Yoshiyuki Obuchi[j], John Pazder[f], Vladimir Reshetov[f], Reed Riddle[a], Sakae Saito[j], Roger Smith[a], Ji Man Sohn[b], Fumihiro Uraguchi[j], Tomonori Usuda[j], Eric Wang[b], Lianqi Wang[g], Jason Weiss[j] and Robert Wooff[f]

[a]Caltech Optical Observatories,1200 E California Blvd., Pasadena, CA 91125;
[b]Department of Physics and Astronomy, University of California, Los Angeles, CA 90095-1547;
[c]Dunlap Institute for Astronomy & Astrophysics, University of Toronto, ON, Canada, M5S 3H4;
[d]Department of Astronomy & Astrophysics, University of Toronto, ON, Canada, M5S 3H4;
[e]Lawrence Livermore National Laboratory, 7000 East Ave., M/S L-210, Livermore, CA 94550;
[f]Herzberg Institute of Astrophysics (HIA), National Research Council Canada, 5071 W Saanich Rd, Victoria, V9E 2E7;
[g]Thirty Meter Telescope Observatory Corporation, 1111 S. Arroyo Pkwy, #200, Pasadena, CA 91105;
[h]University of California Observatories, CfAO, University of California, 1156 High St., Santa Cruz, CA 95064;
[i]Dominion Astrophysical Observatory, National Research Council Canada, W Saanich Rd, Victoria, V9E 2E7;
[j]National Astronomical Observatory of Japan, 2-21-1 Osawa, Mitaka, Tokyo, 181-8588 Japan;
[k]Nanjing Institute of Astronomical Optics and Technology, Chinese Academy of Sciences, 188 Bancang St, Nanjing, Jiangsu, China 210042.

## ABSTRACT

We present an overview of the design of IRIS, an infrared (0.84 - 2.4 micron) integral field spectrograph and imaging camera for the Thirty Meter Telescope (TMT). With extremely low wavefront error (<30 nm) and on-board wavefront sensors, IRIS will take advantage of the high angular resolution of the narrow field infrared adaptive optics system (NFIRAOS) to dissect the sky at the diffraction limit of the 30-meter aperture. With a primary spectral resolution of 4000 and spatial sampling starting at 4 milliarcseconds, the instrument will create an unparalleled ability to explore high redshift galaxies, the Galactic center, star forming regions and virtually any astrophysical object. This paper summarizes the entire design and basic capabilities. Among the design innovations is the combination of lenslet and slicer integral field units, new 4Kx4k detectors, extremely precise atmospheric dispersion correction, infrared wavefront sensors, and a very large vacuum cryogenic system.

## 1. INTRODUCTION

The InfraRed Imaging Spectrograph (IRIS, Larkin et al.[1]) is being developed as a first light, facility instrument for the Thirty Meter Telescope (TMT, Sanders et al.[2]) being constructed on the summit of Mauna Kea in Hawaii. IRIS utilizes the advanced adaptive optics system NFIRAOS (Herriot et al.[3]) and has integrated on-instrument wavefront sensors (OIWFS, Dunn et al.[4]) to achieve diffraction limited resolutions at wavelengths longer than 1 μm. IRIS combines a powerful "wide field" imager and an integral field spectrograph both covering wavelengths from 0.84 μm to 2.4 μm. Both the imager and spectrograph are built around the latest 4K by 4K HgCdTe detectors from Teledyne (Hawaii 4RG). The imager has 4 milliarcsecond (mas) pixels and is baselined with one detector array generating a fully AO-corrected field of view of 16.4 x 16.4 arcsec$^2$. Designs for a larger imager channel are being investigated that would double or quadruple the covered field. The spectrograph offers four plate scales ranging from 4mas to 50 mas and can take up to 10,000 spectra simultaneously in a filled rectangular pattern. A spectral resolution of 4000 is provided for all broad bandpasses and select higher spectral resolutions are also supported. IRIS is a general-purpose instrument working at all distance scales, from our Solar System to the edge of the observable Universe with the very first galaxies.

---

[*] amoore@astro.caltech.edu; phone 1 626 395-8918; fax 1 626 568-1517

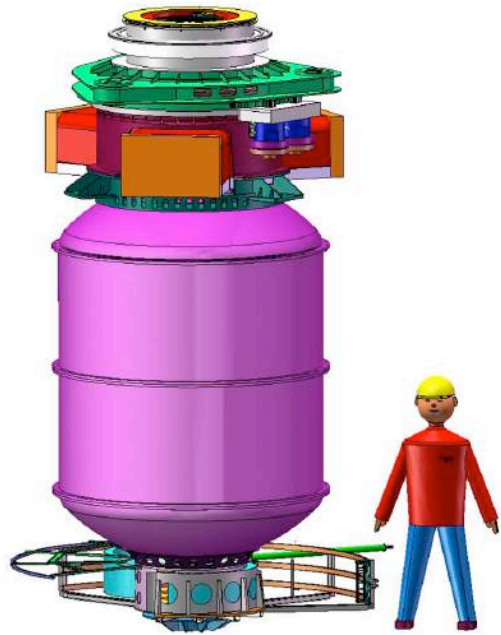

Figure 1: The entire IRIS instrument shown next to a 6-foot tall model human. Light from the AO system NFIRAOS enters from the top through the insulated snout. The entire instrument is suspended from the AO system and rotates in the position shown about its central axis with a fixed gravity vector.

We present the top-level summary of the IRIS capabilities in Table 1, and the performance of the combined NFIRAOS/IRIS system in Table 2. These performance models are based on extensive simulation and include predicted limiting magnitude (Wright et al.[5]), image quality in terms of Strehl ratio and ensquared energies, the level of atmospheric dispersion correction, astrometric accuracy and level of sky background subtraction.

The science fields of view are presented in Table 1. The slicer spectrograph offers a fixed array of 45x88 spatial elements in its rectangular field of view, and the lenslet array spectrograph has a normal mode of 112x128 spatial elements and a secondary field of view of 16x128 spatial elements to increase the number of available spectral elements. The imager mode can be used in parallel and independently with any of the spectrograph modes.

We begin with a summary of the science objectives, followed by a description of the TMT AO system, NFIRAOS, and expected performance. The instrument calibration system is summarized. A technical description of the IRIS instrument is provided with suitable design rationale. Lastly, we present the status that includes a series of trade studies addressing a possible doubling or quadrupling of the imager field of view, investigating the optimum grating technology, ADC design and high contrast imaging and spectroscopy.

**1.1 Scientific Objectives**

In the new era ushered in by TMT, science will evolve extremely rapidly. Judging from our vantage point in 2014, it is difficult to predict what TMT's most compelling discoveries will be. However, if we extrapolate from the most intriguing science of today, we see a clear and well-defined set of primary capabilities that will push TMT to its limits and expand the frontiers of astronomy. Details of individual science cases are discussed in Barton et al.[6] and Do et al.[7] and sensitivity and performance estimates are given in Wright et al.[5]

**Table 1:** A top-level summary of IRIS capability

| Capability mode | Spatial sampling (mas) | Field of View (arcsec) | Resolution ($l/dl$) | Min/Max wavelength (m$m$) | Bandpass | Coronagraph |
|---|---|---|---|---|---|---|
| Imager | 4 mas | 16.4 x 16.4 Possible expansion to 32.8x16.4 or 32.8x32.8 | Set by filter | 0.84-2.4 | 37 filters Variety of bandpasses | Planning mask mounted in OIWFS |
| Slicer IFS 88x45 Spaxels | 50 mas 25 mas | 4.4 x 2.25 2.2 x 1.125 | 4,000, 8,000 4,000, 8,000 | 0.84-2.4 0.84-2.4 | 20%,10% 20%,10% | Planning mask mounted in OIWFS |
| Lenslet IFS 112x128 Spaxels | 9 mas 4 mas | 1.01 x 1.15 0.45 x 0.51 | 4,000 4,000 | 0.84-2.4 0.84-2.4 | 5% 5% | Planning mask mounted in OIWFS |
| Lenslet IFS 16x128 Spaxels | 9 mas 4 mas | 0.144 x 1.15 0.064 x 0.51 | 8,000-10,000 8,000-10,000 | 0.84-2.4 0.84-2.4 | 20% 20% | Planning mask mounted in OIWFS |

**Table 2:** Top-level performance characteristics of the IRIS instrument

| Performance category | Value | Comment |
|---|---|---|
| Expected Strehl ratio for greater than 50% of sky | J band: 0.41 H band: 0.60 K band: 0.75 | For on-axis object. Relative Strehl ratio variations of 1.5-2.5% across entire IRIS field on account of multi-conjugate correction. |
| Airy ring size | J band: 21 mas H band: 28 mas K band: 37 mas | Diameter (FWHM) |
| Ensquared energy | J band: 0.35 - 0.57 H band: 0.50 - 0.66 K band: 0.62 - 0.72 | Over 16.4 x 16.4 arcsec$^2$ imager field. Energy in box with diameter of PSF FWHM Uncertainty originates from different conversions between WFE and EE. |
| Astrometric accuracy | Relative precision: 10 mas Relative accuracy: 30 mas Absolute accuracy: 2-4 mas | Relies on multiple visits to a particular field and a variety of reference fields. |
| Limiting magnitude (Imager) | J band: 27.8 H band: 27.3 K band: 26.9 | Point source sensitivity. Five hour integration, S/N=100, 2l/D aperture. AB magnitude. |
| Limiting magnitude (Spectrograph) | J band: 25.8 H band: 26.0 K band: 25.2 | Point source sensitivity for 4 mas pixel scale. Other scales are significantly more sensitive. Five hour integration, S/N=10, 2l/D aperture. AB magnitude. |

A large international science team (24 members) from six countries is continually working to refine the capabilities of IRIS and motivate its construction. For example, there is an IRIS astrometry team that includes a subset of science and technical members that are dedicated to investigate the astrometric requirements (i.e., 30 mas relative and 4mas absolute astrometry) and the global error budget from the telescope, NFIRAOS, and IRIS system. Our science team has investigated the many science cases that TMT and IRIS will offer to a range of astronomical fields. These include but are not limited to the first galaxies to form, galaxies in the early universe, gravitational lensing, nearby galaxies, supermassive black holes, Galactic Center, microlensing, extrasolar planets, and the Solar System.

## 2. TMT & NFIRAOS

### 2.1 Observatory Environment

TMT will be an altitude over azimuth telescope with a segmented primary mirror spanning 30m in diameter. The telescope supports science instruments on two fixed elevation Nasmyth platforms as shown in Figure 1(a). The telescope tertiary mirror is steerable and capable of feeding any one of the instruments. Furthermore the tertiary mirror is capable of switching between instruments with very little time overhead. Figure 1(a) shows the optical path from the tertiary mirror feeding the TMT Adaptive optics system called NFIRAOS. The focal ratio of the delivered beam of TMT is F/15.

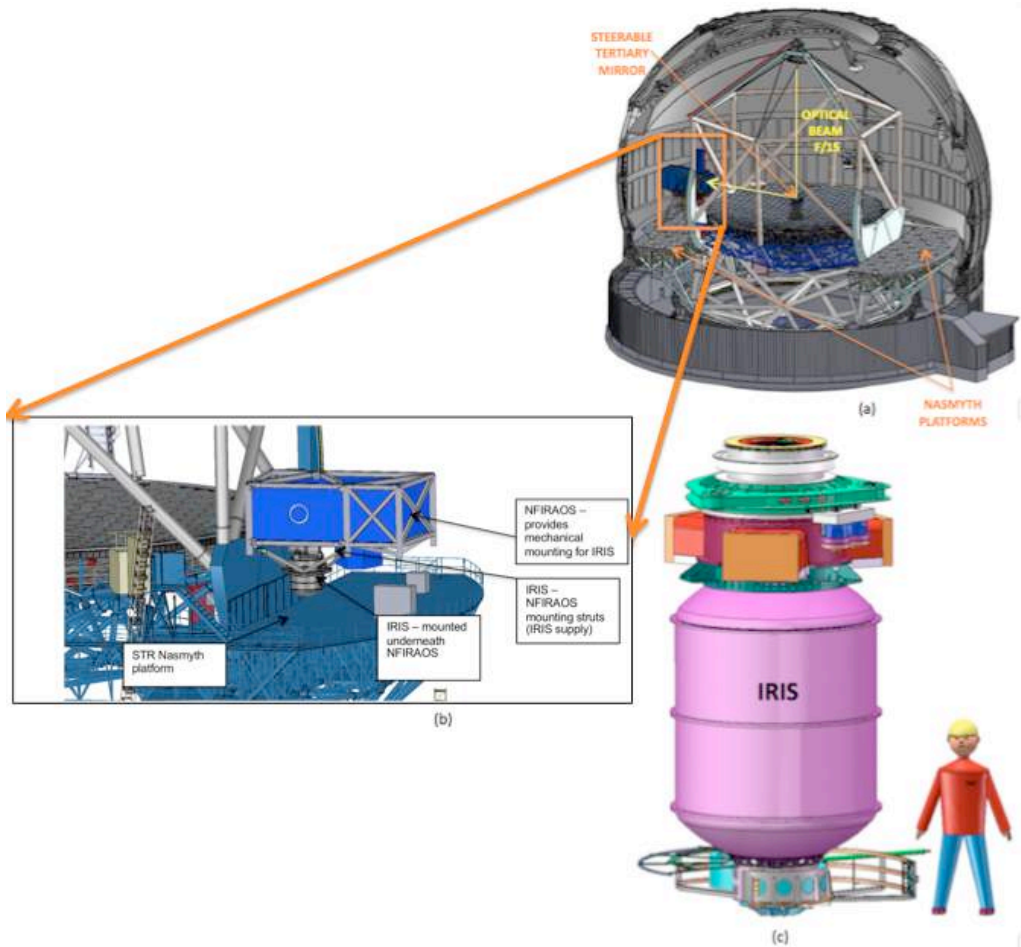

**Figure 1: (a)** The Thirty Meter Telescope shown with cut-away dome housing. Instruments are located on one of two Nasmyth platforms. Light from the secondary mirror is diverted to the Adaptive Optics system (NFIRAOS) by a steerable tertiary mirror. **(b)** NFIRAOS is enclosed inside a cooled chamber to reduce thermal background. Three AO-fed instruments are directly supported by the NFIRAOS structure with the corresponding volume allocations shown. IRIS is located on the lower port that is gravitationally invariant. (c) IRIS is a cryogenic instrument roughly 3.5m high and 2m wide. Cables are fed to the instrument from the lower cable wrap. The instrument rotates to counteract the field rotation of the altitude over azimuth telescope.

## 2.2 NFIRAOS & Sky Coverage

The Narrow Field Infrared Adaptive Optics System (NFIRAOS) is the TMT Observatory's initial facility AO system. It is a multi-conjugate AO system feeding science light from 0.8 to 2.5 microns wavelength to near-IR client instruments. NFIRAOS is located at one of the two Nasmyth platforms of TMT, and supports up to three client instruments on top, bottom and side ports as shown by the cylinders in Figure 1(b). The overall size of the NFIRAOS enclosure is 9 x 6 x 3 $m^3$. NFIRAOS preserves the TMT focal ratio of F/15. It provides an optical throughput of at least 80% in the near-IR, and is cooled to -30 C to reduce thermal emission to below 15% of the contributions from the sky and the telescope optics. The optical distortion is on the order of 0.01% to enable high precision astrometry.

NFIRAOS can utilize a single natural guide star when available, but can also operate with constellation of artificial sodium laser guide stars (LGS) to produce tomographic measurements and multi-conjugate corrections of the turbulence. In this mode, NFIRAOS provides a sky coverage of 50% at the Galactic pole with the near-diffraction limited performance (<190-200 nm RMS wavefront error over a 30 arc sec diameter field), as well as a uniform and stable PSF for high precision astrometry and photometry. Figure 2 plots the median performance (i.e., 50% sky coverage) of NFIRAOS over the full sky with median seeing at the TMT site. The recent addition of a LQG (Linear Quadratic Gaussian) controller we expect will improve the rejection of vibration and hence improve the sky coverage presented in Figure 2 (Wang et al[8]).

IRIS is one of three AO-fed instruments connected to the NFIRAOS enclosure. The arrangement and permissible instrument volume are shown in Figure 1(b). IRIS is allocated the lower port that is gravitationally invariant. This substantially eases the design requirements for such a large cryogenic instrument working at the diffraction limit of the telescope. The choice of altitude over azimuth results in a field rotation at the delivered focal plane as the telescope tracks across the sky. Given the necessary field sizes and resulting size of de-rotating optics, NFIRAOS itself does not provide field de-rotation. Instead the instruments are physically rotated about their optical axis. For IRIS this is just a simple rotation about its long axis without any variation in gravity.

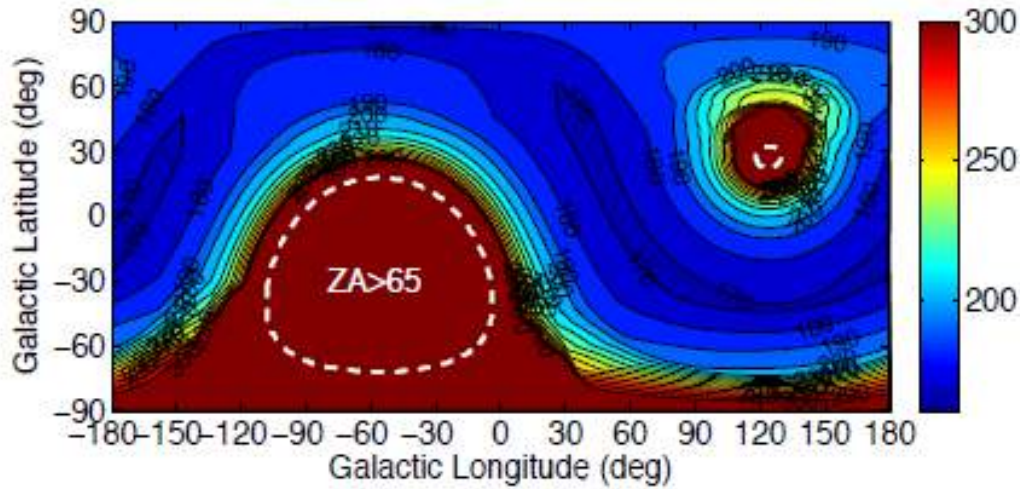

**Figure 2:** RMS wavefront error (nm) at 50% sky coverage and median seeing, averaged over a 17x17" FoV for an object at transit. The dashed lines indicate the TMT zenith angle limit of 65 degrees.

## 2.3 NCSU Calibration Unit

A common calibration unit (Moon et al[9]) is located at the front of NFIRAOS that contains all necessary calibration illumination for NFIRAOS and its client instruments. The NCSU calibration unit will provide the IRIS instrument arc lamps, flat fields and precision grid maps necessary for calibration of the various IRIS modes. It is possible a lenslet calibration system will be located within the IRIS dewar itself.

## 3. IRIS TECHNICAL DESIGN

The IRIS instrument, shown in Figure 1(c) and Figure 3 is 3.5m high and approximately 2m in diameter. IRIS consists of four general components as labeled in Figure 3: the mechanical and thermal interface to the NFIRAOS system, the On-

Instrument Wavefront Sensor (OIWFS) enclosure that is maintained at -30 C, the large vacuum Science Dewar maintained at cryogenic temperatures and the cable wrap. The science imager and integral field spectrograph channels are located inside a vacuum vessel, colored purple in the figure, and held at temperatures ranging from 77K-120K. The vessel is large and will require a long time to pump and thermally cycle, possibly of the order of 4 weeks. It is therefore imperative to deliver IRIS fully tested so as not to require any thermal cycling once commissioned. IRIS is a fully cryogenic diffraction limited imager and integral field spectrograph with four plate scales ranging from 4 to 50 milliarcseconds, and at least 20 filters and 12 grating combinations. The design attempts to optimize each of these modes with as many common optics as possible and without requiring a large number of cryogenic mechanisms.

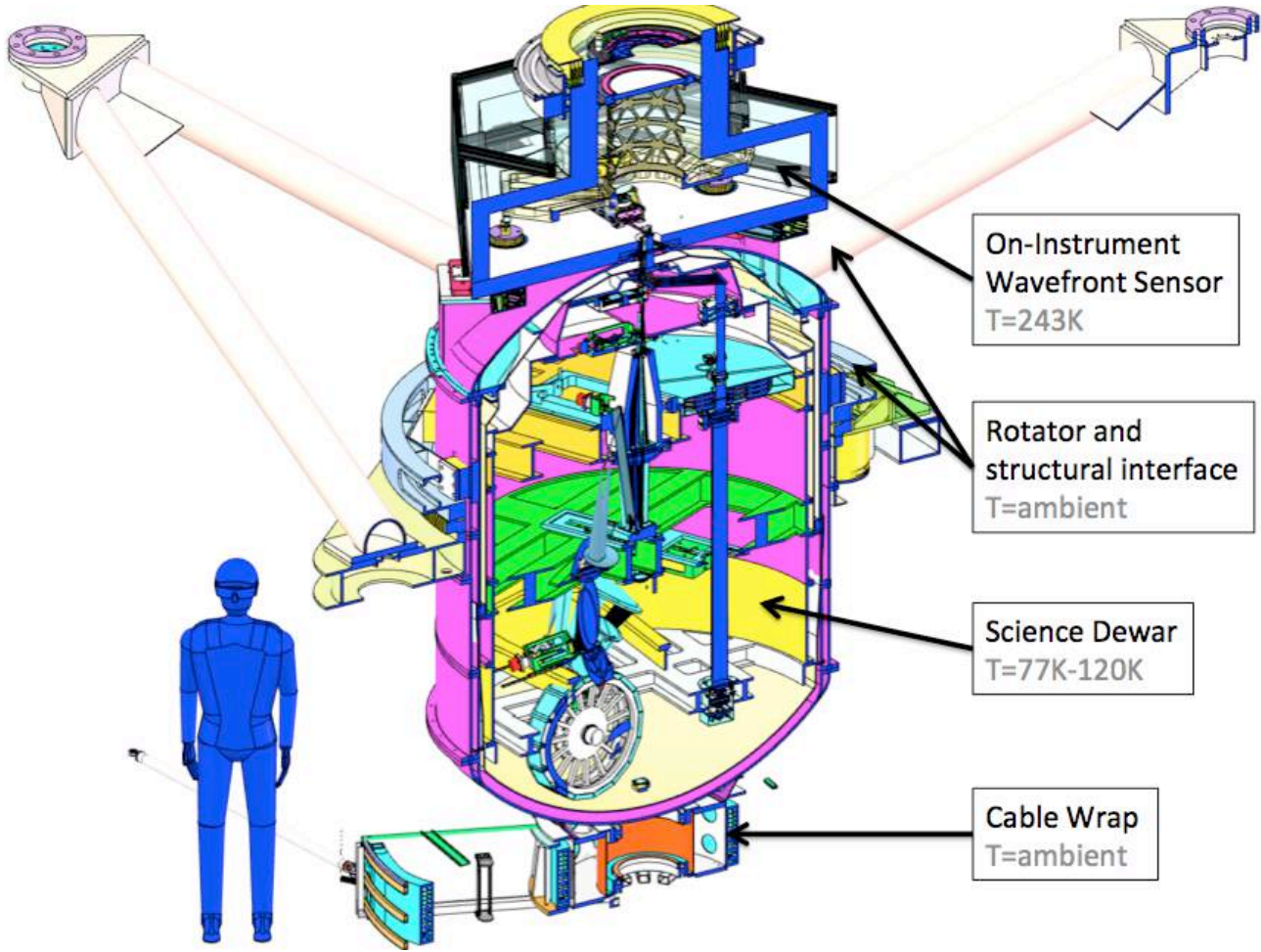

**Figure 3:** *This cut-away of the IRIS instrument shows the major instrument sub-assemblies. Only the vacuum-sealed science dewar is maintained at cryogenic temperatures. The remaining modules are cooled to -30$^o$C, the NFIRAOS enclosure temperature, or held at ambient. The science dewar contains the IRIS imager and IFS channels.*

### 3.1 Rotator and NFIRAOS Interface

The instrument mounts to the lower port of the NFIRAOS instrument and can rotate in place without changing the gravity orientation of the instrument. However, rotating such a large cryogenic instrument with microarcsecond astrometric precision is very challenging. Shown in Figure 3, the instrument rotator is at the top of the dewar and is integrated with the wavefront sensor package. The large rotary bearing provides support for the instrument and is also thermally insulated from the NFIRAOS. A snout extends into the NFIRAOS output port to provide an insulated connection between the -30 C NFIRAOS chamber and the -30 C OIWFS section. In the current design, a cable wrap at the bottom of the instrument allows all of the umbilical connections to wrap around the instrument as it rotates without hindering the field rotation. The mechanical interface between IRIS and NFIRAOS is currently being optimized with the purpose of addressing the strict requirements for earthquake safety (see Dunn et al[4]).

### 3.2 On-Instrument Wavefront Sensor (OIWFS)

The NFIRAOS adaptive optics system will often use laser guide stars and perform a multi-conjugate analysis of the atmosphere. This allows it to correct stellar images over a full 2 arcminute field, but the LGS AO system itself is blind to several of its own corrective modes. In particular, field distortion, tip/tilt and focus must be determined independently using real stars and natural guide star sensors. For full correction of these effects, three stars must be monitored in the corrected field of view. Given the density of stars on the sky, achieving full correction over at least half of the sky (or equivalently for half of anyone's targets of interest) requires using very faint stars where AO correction is needed to improve their contrast against the background light. As a result, IRIS is designed with three pick-off arms which patrol the AO corrected field of view of NFIRAOS and which will use infrared wavefront sensors to gain from the high Strehl ratios in the infrared. To further improve the sensitivity, the sensors will integrate over a very wide wavelength range (0.84-2.3 microns). The sensors are cryogenic infrared detectors probably cooled to liquid nitrogen temperatures, but the optical components including probe arms must be cooled to roughly -30 C to reduce the thermal background.

Since the OIWFS serves as the direct measurement of stars on the sky, its performance is one of the dominant factors in the astrometric performance of the system. In addition, there is differential atmospheric dispersion for the LGS wavefront sensors, OIWFS, and science cameras, which is a function of wavelength and position on sky. This means that positions of the guide stars and the science target(s) are moving differentially in real time with respect to each other. All of these effects are generating "breathing" modes in the AO system that can impact relative astrometric accuracy and precision measurements on the science camera. As a result, the OIWFS probe arms must move very accurately (typical rms of better than 4 μm) and must track well as guide stars transition across pixel boundaries.

Given this requirement a prototype probe arm was assembled and fully tested at the Herzberg Institute of Astrophysics (HIA). The probe arm, shown in Figure 4 under test, successfully met all requirements both at laboratory and -30 C temperatures. For an in depth description of the OIWFS see Dunn et al.[4].

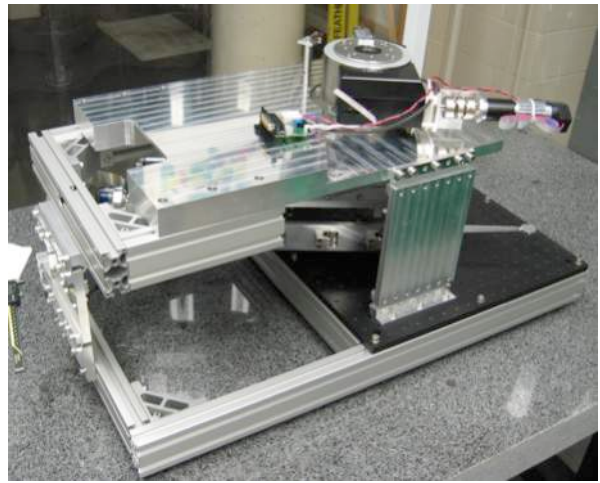

Figure 4: Testing of the prototype low order wavefront sensor probe arm at the Herzberg Institute of Astrophysics

### 3.3 Diffraction limited Imager

The imager is designed to sample the diffraction limit of TMT at all IRIS wavelengths beyond 1 micron with a goal of 30 nm of rms wavefront error. With the expectation of detectors with 4096 pixels on a side (16 megapixels), and sampling with 4 mas per pixel, we have designed optics to provide this image quality over a 16.4x16.4 arcsecond field. We looked at many configurations using reflective and refractive elements and found that a refractive option based on apochromatic triplets best met the requirements. The optical system consists of a collimator and camera both with a $BaF_2$-Fused Silica-ZnSe apochromatic triplet and a single $BaF_2$ lens near the focus. The rms wavefront error of the system is less than 22 nm with ideal optical parameters. A sensitivity analysis shows that a reasonable amount of errors in fabrication and alignment will give the rms wavefront error of less than 30 nm in 90 % of all cases. We find that given expected mechanism repeatability, distortion errors are well below 100 micro-arcseconds. For more information on the imager see Suzuki et al[10].

## 3.4 Hybrid Spectrograph and Slicing Technique

Perhaps the biggest decision in an integral field spectrograph is the method of "slicing" the two dimensional field such that the third spectral dimension can be formatted onto a two dimensional detector array. In any technique there is a strong tension between field of view and the number of spectral channels given the limited number of detector pixels. With TMT this is particularly challenging since in some cases the diffraction limited core (sometimes under 10mas in diameter) can be 100 times smaller than the halo of the point spread function (PSF~1 arcsecond). So to properly sample even a single PSF could require 100x100 spatial elements with relatively short spectra. During the IRIS feasibility phase (2005-2006) a trade study was performed between a "lenslet-only" and "image slicer-only" spectrograph for IRIS. The conclusion reached was that each technology had strong advantages and disadvantages, depending upon the desired plate scale and science case. Table 3 summarizes the conclusion of this trade study. Both designs used approximately the same number of spaxels (a spatial/spectral location within the data cube) as in the present design (112x128 lenslets and 88 slicers with 45 spatial pixels each, respectively) and had comparable throughput.

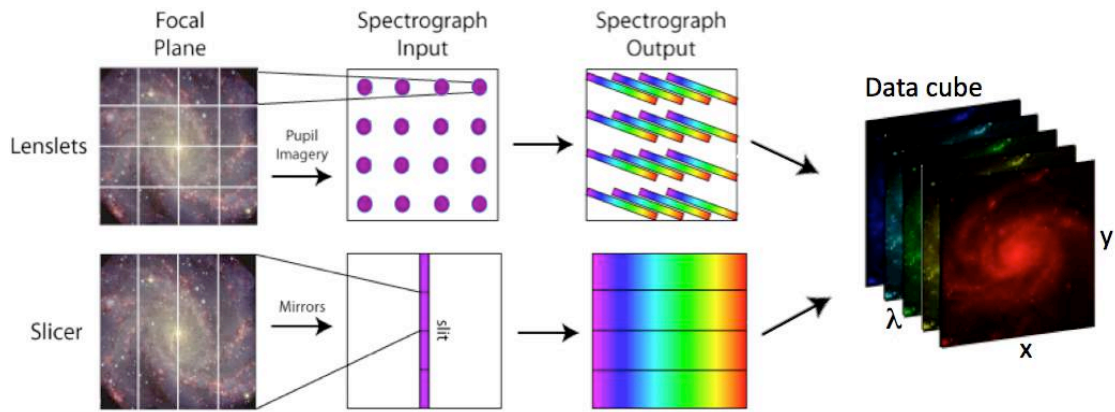

**Figure 5:** Schematic representation of a lenslet based and slicer based integral field spectrograph. Figure is for clarification (4x4 sampling for lenslet array and 4 slicers only) and does not represent IRIS implementation that has many more spatial elements. With lenslets, a grid of small (350 micron pitch) lens are placed in a focal plane and concentrate the light into an array of tiny pupil images each representing the light from one pixel on the sky. A relatively standard spectrograph then disperses all of the pupil images into an interleaved set of spectra. For a slicer, a set of long parallel mirrors are placed in the focal plane each tilted by a small relative amount compared to its neighbor. A second set of mirrors then redirect these slivers of light into a long line of light that can be fed into a spectrograph much like a traditional long slit spectrograph. The final spectra from both spectrographs can be extracted and reorganized into a data cube containing spatial and spectral information from a contiguous patch of sky.

The finest spaxel scale of 4mas samples the diffraction limit of the telescope at a wavelength of 1um. The IFS must have an extremely low residual wavefront error (WFE<30nm) for this to be achieved in practice. A lenslet-style spectrograph is superior at the finest scale since the lenslet itself samples the image plane after a minimal set of reimaging optics. The lenslet array concentrates each focal plane location into a small and well-separated dot of light and prevents later optical elements (gratings, TMA's, etc…) from affecting the image quality. In addition, it is easier to expand a lenslet-style spectrograph to many spatial locations allowing for a field of view sufficient to image a large fraction of a PSF halo even at the smallest scales. At coarse scales, as the field of view starts to exceed 1-2 arcseconds, science cases prefer to expand spectral coverage to the full 20% bandpasses for simultaneous comparisons of spectral lines under the same conditions. Here a slicer is superior since the lenslet would be wasting most of its spaxels on sky and cannot be reformatted to satisfy the larger bandpass. The two intermediate scales are more ambiguous but the FOV benefit of the lenslet aids with the 9 mas scale while the slicer makes better use of its pixels in the 25 mas scale. What Table 3 does not show is that either technique struggles to handle the full range of magnifications simply due to the factor of 12.5 difference in final focal ratios between 4 and 50 mas. In general this means different plate scales produce beams of light 12.5 times different in size on any common optic. So any shared optics would typically need to be large for the coarse scales and have exquisite shapes for the narrow scales.

A surprising outcome of the competitive feasibility study was that an F/4 camera system would work well for the two fine plate scales with a lenslet array, and the two coarse scales with the slicer array. So a hybrid design is not only feasible, but is actually cheaper and perhaps even simpler in some ways than a lenslet or slicer design that attempted to handle all four plate scales. And the hybrid design uses each type of spatial sampler (lenslet or slicer) when it is best suited. So our team combines instrument expertise in these two spectrograph types and is uniquely suited for its design and fabrication. For details of the lenslet and IFS channels see Moore et al[11].

**Table 3:** Results of image slicer-only versus lenslet-only technology for IRIS. Led to the choice of the hybrid design. Rows shown in gray are for magnifications where the particular technique has some problems or challenges.

**Hypothetical Parameters for Slicer-only Spectrograph (shaded=poor, white=good performance)**

| Scale (mas) | FOV (") | Bandwidth | Comment |
| --- | --- | --- | --- |
| 4 | 0.18x0.35 | 20% | Poor wavefront quality, very small field |
| 9 | 0.40x0.79 | 20% | Image acceptable, small field |
| 25 | 1.13x2.20 | 20% | Field acceptable, good use of pixels for bandwidth |
| 50 | 2.25x4.40 | 20% | Good FOV, good bandwidth |

**Hypothetical Parameters for Lenslet-only Spectrograph**

| Scale (mas) | FOV (") | Bandwidth | Comment |
| --- | --- | --- | --- |
| 4 | 0.45x0.51 | 5% | Reasonable field size, excellent image quality |
| 9 | 1.01x1.15 | 5% | Good FOV, excellent images |
| 25 | 2.80x3.20 | 5% | Most cases would prefer more bandwidth vs. field |
| 50 | 5.60x6.40 | 5% | FOV excessive, strongly prefer more bandwidth |

### 3.5 Atmospheric Dispersion Correction

The angular resolution of IRIS turns some traditionally minor inconveniences into major problems. Atmospheric dispersion is a concern at TMT's spatial resolution and is something the team has worked hard to understand; an analysis is presented in detail in Phillips et al[12]. A star's position, even across a 20% bandpass filter, can vary by many resolution elements from the blue end of the bandpass to the red end. At 1 micron (Y-band), the diffraction limited core has a FWHM of roughly 8 mas, but at zenith angles around 20 degrees, the differential dispersion is already more than 30 mas or roughly 4 resolution elements. The direction of the dispersion is also always oriented along the elevation axis, so it sweeps out an arc over time compared to right ascension and declination. So even in an integral field spectrograph, where independent images are maintained at fine wavelength steps, a star's position cannot be held fixed in more than one wavelength slice at a time without dispersion compensation. To perform astrometry at the tens-of-microarcseconds level, real time and highly accurate dispersion correction is needed. To achieve this, we have looked at hundreds of glass combinations to find ones that work well over the full wavelength range (0.8 – 2.4 µm), both for the OIWFS probes and for the science instruments. For reasons discussed in Phillips et al[12], we have selected crossed Amici prisms to correct for the atmospheric dispersion. The counter-rotating prism pairs can produce a stellar image with less than 1 mas of residual motion across any of our bandpasses, and with knowledge at a significantly higher level. We've also investigated dispersion issues of field distortion introduced by the atmosphere, centroid shifts that can occur due to the color of the tip/tilt stars, and the effect of uncertain atmospheric conditions (pressure and temperature) at the telescope at the time of an exposure.

### 3.6 Mechanical Design

Thermal and mechanical stability are crucial for the performance of the IRIS instrument and dramatically affect the ability to reduce and analyze the complex data products. In all aspects of the design, we have started with existing reliable solutions, especially in the area of cryogenic mechanisms and optical mounts. The vacuum vessel is approximately 2 meters in diameter and 3.5 meters long. This is comparable to the diameter of the MOSFIRE instrument (McLean et al[13]), which was a similar collaborative project between UCLA, Caltech, UCSC and the Keck Observatory. Figure 6 shows the vacuum chamber in comparison to the MOSFIRE instrument.

As with any cryogenic instrument, especially at the size scale of IRIS, mechanisms should be minimized and must be extremely robust (typically with 10-year demonstrated life). For the science dewar, where mechanisms will likely operate between 77 and 120 Kelvin, there are 16 cryogenic mechanisms in the current design and most are baselined on former OSIRIS (Larkin et al[14]), MOSFIRE and GPI (Gemini Planet Imager, MacIntosh et al[15]) designs. But there are also several new challenging mechanisms like the large grating turret that must repeat to only a few microns when moving twelve large gratings. Like all of our most challenging elements, we prototyped this grating turret in 2013 and have a working cryogenic mechanism based on some of our largest previous wheel mechanisms, but with some entirely new solutions. As assemblies for the final instrument are developed these will require extensive testing before being installed in the dewar. A set of large test chambers have been designed (one each at UCLA, Caltech and NAOJ) to test systems in parallel without the long cool down times of the full system.

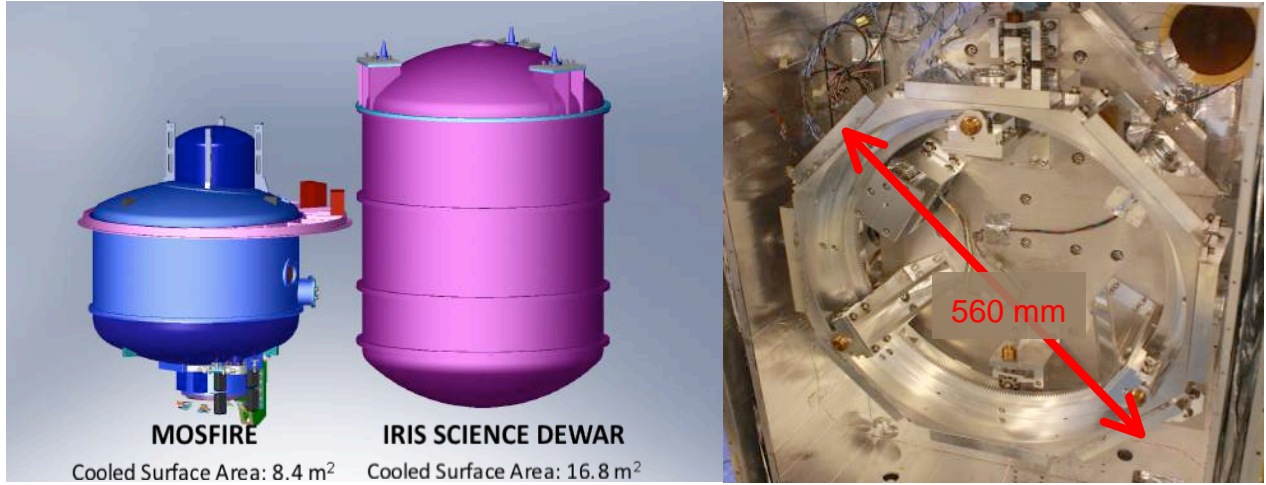

**Figure 6** (left)**:** *The MOSFIRE spectrograph has recently been delivered at the Keck telescope and was cooperatively built by Caltech and UCLA in a very similar fashion as proposed for IRIS. The large cryogenic dewar also serves as a good comparison for the proposed IRIS vacuum chamber. Domed end caps and several separate sections provide access to the internal components.*

**Figure 7** (right)**:** *The IRIS prototype grating turret over half a meter in diameter. It can move cryogenically to any position in less than 30 seconds and each grating can be positioned repeatedly within a few microns.*

### 3.7 Gratings

In most spectrographs the grating is one of the largest causes of throughput loss. Facet quality is critical in a ruled grating and can prove challenging in infrared instruments which often have coarse facets. Volume phase holographic (VPH) gratings (Barden et al[16]) offer a new promising method of producing moderate spectral dispersion with high throughput. But as a relatively new technology they have a limited history especially in cryogenic environments in the infrared. Furthermore, we had two concerns with the application of VPH gratings for IRIS. Firstly, the required line density for the R=4000 gratings was considered to be on the low side from a manufacturing perspective by the vendors (<200 lines/mm is required for the broadband H and K gratings). Secondly, the IRIS lenslet in full field mode and the slicer IFS channels re-image to the detector at a range of angles corresponding to a varying angle of incidence at the grating *in the spectral direction*. This angle is significantly off-Bragg leading to a throughput concern especially in the broadband gratings. The narrow field lenslet mode is much more insensitive to this effect.

Given the critical nature of the gratings in IRIS we began a prototyping project to test head-to-head diffraction ruled gratings and VPH gratings from two companies each. Representative gratings were made in 2013 for the low and high resolution J and H bands. Throughput and polarization testing was completed at the Dunlap Institute for Astronomy and Astrophysics during 2013 and early 2014. Preliminary results of this trade study are presented at this meeting (Chen et al[17], Meyer et al[18]).

### 3.8 Detectors

IRIS works at very fine sampling of the sky and with relatively high spectral dispersion. Our goals also include observing the faintest sources ever attempted. In many science cases, our backgrounds and sensitivities are directly proportional to the detector performance. In the 1-2.5 micron regime, HgCdTe detectors (Blank et al[19]) remain the unquestioned choice due to their extremely low dark currents and read noises, while maintaining high quantum efficiency (not quite as high as InSb). In principle a 2x2 array of existing Hawaii-2RG (4 million pixels each) would match the current optical designs, but the gaps between the arrays would cause problems given the staggered nature of the spectra in the focal plane. The new Hawaii-4RG provides 16 million pixels, comparable pixel performance plus several improvements in reference pixel access and general operation.

## 4. SUMMARY

We have developed a conceptual design for a near-infrared integral field spectrograph and imager that will work at the diffraction limit of the Thirty Meter Telescope. We have working designs for all components and subsystems of the imager and spectrograph and are midway through the preliminary design phase. The instrument builds on the heritage of several previous instruments including the Keck instruments OSIRIS and MOSFIRE.

### 4.1 Updates since concept design review (Dec 2011)

An 18-month long prototyping phase immediately followed the completion of the concept design. During this phase the following was achieved:

1. The large cryogenic grating turret was assembled and successfully cold tested at UCLA;

2. An OIWFS probe arm was assembled and successfully tested at HIA. Both mechanisms require low temperature operation with high repeatability and both met the requirements;

3. The material Spinel, baselined for the IRIS ADC prisms during the concept design, was investigated. Primarily produced in large sizes as a bulletproof glass, its optical properties were still somewhat uncertain, especially its scattering properties. Spinel material blank was acquired by HIA during the prototyping phase and tested, however, the homogeneity and scattering performance was not satisfactory for application to the IRIS ADC prisms. Alternative glass combinations are under investigation.

The project began the preliminary design phase in April 2013. A series of focused trade studies are currently underway and we present the latest progress here:

1. The size and division of the focal plane delivered by NFIRAOS is under investigation. Imager designs that permit a doubling or quadrupling of the imager field of view are being developed. Designs that permit the IFS and imager fields to be as close as possible on the sky are preferred;

2. A "global" ADC is being investigated that notionally would be housed in the collimated space of NFIRAOS. If the performance is acceptable this method may provide adequate correction of the entire 2 arcmin field delivered to the instrument and not just the science field;

3. A throughput performance comparison between diffraction ruled gratings and equivalent VPH gratings is underway with preliminary results presented at this meeting;

4. A study into the application of IRIS to the science of high contrast exoplanet imaging and, in particular, spectroscopy has been initiated. This science case may require the addition of hardware components within IRIS and NFIRAOS at a modest increase in cost for a large return in scientific capability.

## 5. ACKNOWLEDGMENTS

The TMT Project gratefully acknowledges the support of the TMT collaborating institutions. They are the Association of Canadian Universities for Research in Astronomy (ACURA), the California Institute of Technology, the University of California, the National Astronomical Observatory of Japan, the National Astronomical Observatories of China and their consortium partners, and the Department of Science and Technology of India and their supported institutes. This work was supported as well by the Gordon and Betty Moore Foundation, the Canada Foundation for Innovation, the Ontario Ministry of Research and Innovation, the National Research Council of Canada, the Natural Sciences and Engineering Research Council of Canada, the British Columbia Knowledge Development Fund, the Association of Universities for Research in Astronomy (AURA), the U.S. National Science Foundation and the National Institutes of Natural Sciences of Japan.